\documentstyle[twocolumn,aps,float,psfig]{revtex}
\begin{document}
\tighten
\draft
\newcommand{\ds}{\displaystyle}
\newcommand{\be}{\begin{equation}}
\newcommand{\en}{\end{equation}}
\newcommand{\bea}{\begin{eqnarray}}
\newcommand{\ena}{\end{eqnarray}}
\title{Scalar fields in an anisotropic closed universe.}
\author{Mauricio Cataldo$^{\,\,a \, b}$
 {\thanks{E-mail address: mcataldo@alihuen.ciencias.ubiobio.cl}}
, Sergio del Campo $^{c}$
 \thanks{E-mail address: sdelcamp@ucv.cl}}
\address{$^a$
Departamento de F\'\i sica, Facultad de Ciencias,
Universidad del B\'\i o-B\'\i o, Avda. Collao 1202, Casilla 5-C, Concepci\'on,
Chile.
\\ $^b$Departamento de F\'\i sica, Facultad de Ciencia, Universidad de
Santiago de Chile, Avda. Ecuador 3493, Casilla 307,  Santiago, Chile.
\\
$^c$ Instituto de F\'\i sica, Facultad de Ciencias B\'asicas y matem\'aticas,
Universidad Cat\'olica de Valpara\'\i so,
Avenida Brasil 2950, Valpara\'\i so, Chile.
}

\maketitle

\begin{abstract}

{\bf {Abstract:}} We study in this article a class of homogeneous,
but anisotropic cosmological models in which shear viscosity is
included. Within the matter content we consider  a component (the
quintessence component) determined by the barotropic equations of
state, $p=\alpha \rho$, with $\alpha < 0$. We establish conditions
under which a closed axisymmetrical cosmological model may look
flat al low redshift. \vspace{0.5cm}

PACS number(s): {98.80.Cq, 95.35.+d,97.10.Fy,98.80.Hw }

\end{abstract}

\smallskip\

\section{Introduction}

Current observations of luminosity-redshift relations of type Ia
supernovas{\cite{PeSchRi}} and measurements of the anisotropy
cosmic background radiation and mass power spectrum{\cite{LiEfCo}}
provide evidence that the total matter density of the universe
coincides with its critical value. This, agrees with the
theoretical arguments derived from inflation{\cite{GuLiAl}}, where
it is suggested that our universe should become soon flat after a
short period of inflation.

Since, astronomical observations give rise to the bound $\Omega_M  \lesssim 0.3$,
in which baryons and Cold Dark Matter are included, we are in front of a problematic situation.
There exist a sort
of "missing energy" that should represent
something like the $70 \%$ of the critical value.

It has been argued that the simplest explication, a cosmological
constant (vacuum energy density) is consistent with these
results{\cite{Peetal}}. Other alternatives have been considered.
For instance, bulk pressure that is significantly negative, i.e.,
$ \alpha \lesssim -1/3$, where $p = \alpha \rho$ is the effective
equation of state, in which $p$ is the pressure and $\rho$ is the
energy density. Here, this sort of matter could correspond to a
network of topological defects{\cite{ViSh}} (such that strings or
walls) or an evolving scalar field (referred as
quintessence){\cite{Caetal}}, $Q(t)$, in which case the pressure
and the energy density become defined by $\ds p_Q =
\frac{1}{2}\dot{Q}-V(Q)$ and $\ds \rho_Q = \frac{1}{2}\dot{Q}^2 +
V(Q)$, respectively. Here, $V(Q)$ represents the scalar potential
associated to the scalar field $Q$ and the overdots specify
derivatives with respect to time.

The main difference between these two sort of models
i.e., the cosmological constant and the scalar field with a negative
pressure, is that the latter is spatially inhomogeneous and thus can cluster
gravitationally, where the former is totally spatial uniform. In this respect,
the fluctuation of the scalar field could have an important effect on large scale
structure of the universe{\cite{Heetal}}.

Since the total energy density equals the critical density, then
the spatial part of the metric is supposed to correspond to a flat
Friedmann-Robertson-Walker (FRW) metric. However, It has been
mentioned that the observations referred above, i.e. those related
to type Ia supernovas,  do not rule out a different type of
geometry{\cite{Ag}}. There, it was advanced that these
measurements allow an open universe in which the cosmological
constant is vanished.

From the theoretical point of view, it seems that quantum field
theory is more consistent on compact spatial surfaces that in
hyperbolic spaces{\cite{WhSc}}. On the other hand, in quantum
cosmology the "birth" of universes have been described under the
assumption that the three-geometry is characterized by a close
spatial surface. In this way, motivated by quantum cosmology and
by the short period of inflation that the universe underwent at
early time in its evolution, we describe in this paper the
conditions under which a closed universe model may look flat al
low redshift. This kind of situation has been considered in the
literature{\cite{KaTo}}. There, a closed universe with $\Omega_0 <
1$ was studied. Here, $\Omega_0 $ represents the density parameter
associated to the total mass of the universe. Openness is obtained
by adding to the matter density texture or tangled strings with
equation of state $p = -\rho/3${\cite{Da}}. Here, the additional
energy density is redshifted as $a^{-2}$, similar to the curvature
term in a closed universe, where $a$ is the scale factor.
Kolb{\cite{Ko}} studied this sort of matter, arising to the
important conclusion that a closed universe may expand forever at
constant speed.

It is natural to assume the geometry at very early epoch more
general than just the isotropic and homogeneous FRW. Although the
universe, on large scale, seems homogeneous and isotropic at
present, there is no observational data that guarantees the
isotropy in an era prior to the recombination. In fact, it is
possible to begin with an anisotropic universe which isotropizes
during its evolution.

In relation with the matter that we could take into account in an
anisotropic background, may have many possible sources. For
instant, populations of collisionless particles, gravitons,
electric, or magnetic fields, or by topological
defects{\cite{Ba1}}.

The anisotropic dynamics can in general encode either relative
velocity effects or dissipative effects or both{\cite{Ba2}}. In
this respect, it is possible to start with an anisotropic universe
that eventually isotropizes at later time in the evolution of the
universe due to dissipative processes involving the matter that it
contains. Also, this kind of model seems to be more appropriated when
adiabatic theory of galaxy formation is considered~\cite{Ze}. Thus,
it seems quite natural to include in this study a matter component with
this kind of property, in a background which in essence is
anisotropic{\cite{Mc}}.

The aim of the present paper is to study a closed anisotropic
cosmological model, with metric corresponding to
Kanstowski-Sachs{\cite{KaSa}}, where the matter content is
composed by an imperfect fluid together with a scalar field whose
equation of state parameter $\alpha$ remains negative during the
evolution of the universe.


\mbox{} \\
\section{The field equations}

We start by considering the effective Einstein Lagrangian given by
\be
\ds {\cal {L}}\,=\,\frac{1}{\kappa}\,R\,
+\,\frac{1}{2}\,(\partial_{\mu}Q)^2\,-\,V(Q)\,+\,{\cal {L}}_{M},
\label{s1}
\en
where, $\kappa = 16 \pi G$, with $G$ the Newton's gravitational
constant, $R$ the scalar curvature, $Q$ the quintessence scalar
field with associated potential $V(Q)$, and ${\cal {L}}_M$
represents the matter Lagrangian density. We assume that the
matter lagrangian density $\cal L_M$   is associated to a fluid
(characterized by the pressure and energy density, $p_M$ and
$\rho_M$, respectively) which  presents a shear viscosity. By
taking a preferred  timelike vector field (four velocity)
$u^{\alpha}$, which satisfies $u^{\alpha} u_{\alpha} = 1$ and it
is a Ricci eigenvector, we can write the  following matter
energy-momentum tensor
\begin{eqnarray}
T_{\alpha \beta}= (\rho_M + p_M) u_{\alpha} u_{\beta} - p_M g_{\alpha \beta} + 2
\eta_M \sigma_{\alpha \beta},
\end{eqnarray}
where  $\eta_M$ and $\sigma_{\alpha \beta}$ are
the shear viscosity (or coefficient of dynamic
viscosity, $\eta_M \geq 0$) and the traceless shear tensor, respectively. The
shear tensor has the form
\begin{eqnarray}
\sigma_{\alpha \beta} = h_{\alpha}^{\gamma} u_{(\gamma ; \delta)}
h^{\delta}_{\beta}- \frac{1}{3} \theta h_{\alpha \beta},
\end{eqnarray}
where $\theta= u^{\alpha}_{\,\, ; \alpha}$ is the scalar expansion
and $h_{\alpha\,\gamma}$ is the projection tensor defined from the
expression
$h_{\alpha\,\beta}\,=\,g_{\alpha\,\beta}-u_{\alpha}\,u_{\beta}$,
with signature for the metric $(+,-,-,-)$.

In this paper we consider a spatially homogeneous background
spacetime of Kantowski-Sachs type, which, as far as it is known,
it is the only spatially homogeneous models that it is not
included in the Bianchi classification, thus we have
\be
\label{metric}
\ds ds^{2} = dt^{2} -  a^2(t) \left (d\theta^{2}+
sin^2(\theta) d\phi^{2} \right) - b^2(t) dr^{2},
\en
where $ a$ and $ b$ are the scale factors which describe the
anisotropy of the model. This sort of metric combines spherical
symmetry with a translational symmetry in the "radial" direction.
The metric ({\ref{metric}}) has been studied by many authors that
have considered different sort of matter components.            
As examples, it has been considered a homogeneous shear-free
cosmological models with an imperfect fluid matter
content~\cite{MiCr}. On the other hand, a energy-effective-action
related to string theory has been studied~\cite{BaDa}. Here, when
the pseudoscalar axion field is time depending only, it reduces to
that of a stiff perfect-fluid cosmology. Also, a scalar field for
a convex positive scalar potential~\cite{BySc} was taken into
account , among others.

Since the metric ({\ref{metric}}) is spatially homogeneous the scalar field $Q$ can only
depend on time, and thus the time-time component of Einstein's field equations is
\be
\ds \left ( \frac{\dot{a}}{a} \right )^2 + \left (\frac{\dot{a}}{a}\right )
\left ( \frac{\dot{b}}{b}\right )
+ \frac{1}{a^2} = \frac{2 \kappa}{3} \left ( \rho_M + \rho_Q \right ),
\en
where, as was mentioned above,  the dots stand for derivatives
with respect to the cosmological time $t$. From the metric
(\ref{metric}), and considering the comoving frame, i.e.,
$u^{\alpha}= \delta^{\alpha}_{0}$, we find that the components of
the shear tensor are given by
\begin{eqnarray}
\ds \sigma_{11} \,=\,\frac{2}{3} \, b^2 \left(\frac{\dot{a}}{a}
- \frac{\dot{b}}{b}  \right), \nonumber \\
\ds \sigma_{22} \,=\,\frac{1}{3} \, a^2  \left(\frac{\dot{b} }{b }
- \frac{\dot{a} }{a }  \right), \\
\sigma_{33}= \frac{1}{3} \, sin^2(\theta)\,a^2  \left( \frac{\dot{b}}{b } -
\frac{\dot{a} }{a } \right) \nonumber .
\end{eqnarray}
Here, $\sigma_{00} = 0$ and $\sigma^{\alpha}_{\,\,\alpha}=0$.
The other components of Einstein's field equations are
\noindent
\bea
\ds   2\,\frac{\ddot{a}}{a} \,+\, \left (\frac{\dot{a}}{a} \right )^2\,
+ \ds \,\left (\frac{1}{a} \right )^2\,\hspace{2.5cm}  \nonumber \\
=\,-\,\kappa(p_M+p_Q)
-\,\frac{4}{3}\,\kappa\,\eta_M\,\left (\frac{\dot{a}}{a}-\frac{\dot{b}}{b} \right ),
\ena
and
\bea
\ds \frac{\ddot{b}}{b} \,+\,\frac{\ddot{a}}{a}\,
+\,\frac{\dot{a}\,\dot{b}}{a\,b}\,=\,-\,\kappa(p_M+p_Q)
+\,\frac{2}{3}\,\kappa\,\eta_M\,\left (\frac{\dot{a}}{a}-\frac{\dot{b}}{b} \right ).
\ena

In order to solve this set of equations, we need to supply this
set with equation of state for the matter content and the scalar
field. We assume that the matter content satisfies the relation
$p_M = \gamma \rho_M$, where $\gamma$ may bee  a time depending
quantity and its (present) value depends on the characteristics of
matter content. In the following we assume that this constant lies
in the range $0 \leq \gamma \leq 1 $, where the extremes
correspond to dust and stiff fluid, respectively. In the same way,
we shall assume that the scalar field $Q$ satisfies a similar
effective equation of state, i.e., $p_Q = \alpha \rho_Q$, where
now the parameter $\alpha$ is assumed to be negative.

In order to have a universe which is closed, but still have a matter density content
corresponding to a flat universe, we impose the following relations:
\be
\label{cond1}
\ds \kappa \,\rho_{_Q}\,=\,a^{-2},
\en
and
\be
\label{cond2}
\ds \eta'\,=\,\eta_{_M}\,+\,\frac{1}{2}\,\frac{\rho_{_Q}}{\overline{\sigma}},
\en
where $\ds \overline{\sigma}\,=\,\left
(\frac{\dot{a}}{a}-\frac{\dot{b}}{b} \right )$ and $\alpha$ has
been chosen to be equal to $-1/3$\cite{CrdeHe}. Under these
conditions the Einstein's field equations become
\be
\label{00}
\ds \left ( \frac{\dot{a}}{a} \right )^2 + \left (\frac{\dot{a}}{a}\right )
\left( \frac{\dot{b}}{b}\right )\,
=\, \frac{2 \kappa}{3}\, \rho_M,
\en
\bea
\label{11}
\ds   2\,\frac{\ddot{a}}{a} \,+\, \left (\frac{\dot{a}}{a} \right )^2\,
 =\,-\,\kappa\,\gamma\,\rho_M
-\,\frac{4}{3}\,\kappa\,\eta'\,\overline{\sigma},
\ena
and
\bea
\label{33}
\ds \frac{\ddot{b}}{b} \,+\,\frac{\ddot{a}}{a}\,
+\,\frac{\dot{a}}{a}\,\frac{\dot{b}}{b}\,=\,-\,\kappa\,\gamma\,\rho_M
+\,\frac{2}{3}\,\kappa\,\eta'\,\overline{\sigma}.
\ena

This set of equations is similar to that of a matter fluid with
shear viscosity  immersed in a background corresponding to a flat
axisymmetric cosmological model.

\mbox{} \\
\section{Solution to the field equations and some consequences}

In the following we will describe solutions to the set of
equations (\ref{00})--(\ref{33}) in the cases in which
$\eta'\,=\,0$, i.e., where there is not generation of entropy and
$\eta'\,\neq \,0$, where exist generation of entropy. For the
former solutions we calculate the angular distance-redshifts
relations (specifically, for the stiff model with $\eta'=0$) which
are compared with its analogous results corresponding to the flat
spacetime, and for the latter, we calculate the generation of
entropy (for the cases in which $\eta' \neq 0$).

\subsection{Cases $\eta'\,=\,0$}

In this case we describe two possible solutions. One of these is
the vacuum Kasner solution in which $p_M = \rho_M = 0$, and a
stiff fluid with equation of state $p_M = \rho_M \neq 0$, i.e.,
$\gamma = 1$. In the former case it is found that   $a(t)=a_0$ and
$b(t)=t$ is possible solution to the field equations. Here, the
scalar field $Q$ remains constant and thus $\rho_Q = const.=V_0$
and $\ds p_Q = -\frac{1}{3}V_0$. On the other hand, $\eta_M$
becomes $\ds \eta_M = \frac{1}{2} \frac{V_0}{t}$. It seems that we
could have  others Kasner solution, such that $a(t)=t^{2/3}$ and
$b(t)=t^{-1/3}$. However, this sort of solution gives rise to a
shear viscosity which is essentially negative, since $\ds \eta =
-\frac{1}{2 \kappa}\frac{1}{t^{1/3}} $. Therefore, we  disregard
this type of solution.

In the latter case, in which $p_M = \rho_M $, i.e., $\gamma = 1$,
we find a possible solution  in which
\be
\label{ab1}
\ds a(t)=t^{n}$, \hspace{1.0cm}  $b(t)=t^{1-2n},
\en
where  $n$ is a positive number. Here, we get
\be
\ds p_M = \rho_M  = \frac{n(2-3n)}{\kappa t^2}
\en
and
\be
\ds \eta_M = \frac{1}{2 \kappa} \frac{1}{(1-3n)} \frac{1}{t^{2n-1}}.
\en
In order to  obtain $\eta > 0$ we demand $n < 1/3$. Here, it is found that
the scalar field growth as
\be
\ds Q(t) = \sqrt(\frac{2}{3 \kappa}) \frac{1}{1-n} t^{1-n},
\en
and its corresponding potential is given by
\be
\ds V(Q) = \left (\frac{2^{2n-1}}{3 \kappa} \right )^{\frac{1}{1-n}} Q^{-\frac{2 n}{1-n}}.
\en
Notice that this potential decreases when $Q$ increases, since $n < 1/3$.

At this moment, we would like to calculate the luminosity distance
, $d_L(z)$, as a function of the redshift $z$. This concept plays
a crucial role in describing the geometry and matter content of
the universe. From the metric (\ref{metric}) we observe that,
light emitted by the an object of luminosity ${\cal L}$ and
located at the coordinate distance $\theta$, at a time $t$ is
received by an observer (assumed located at $\theta = 0$) at the
time $t = t_0$. The time coordinates are related by the
cosmological redshift $z$ in the $\theta$-direction by the
expression, $1 + z = a(t_0)/a(t)\,\equiv \,a_0/a(t)$. The
luminosity flux reaching the observer is $\ds {\cal F} =
\frac{{\cal L}}{4 \pi d_L^2}$, where $d_L$ is the luminosity
distance to the object, given by $d_L(z)\, = \, a_0 \,sin(\theta(z))
\, (1 + z)$.

In order to obtain an explicit expression for the angular size,
let us now consider  an object aligned to the $\phi$-direction and
proper length $l$, so that its "up" and "down" coordinates are
$(\theta, \phi + \delta \phi, 0)$ and $(\theta,\phi, 0)$. The
proper length of the object is obtained by setting $t = const.$ in
the line-element  metric (\ref {metric}), $ds^2 = - l^2 = - a^2(t)
sin^2(\theta) \delta \phi^2$. Thus, the angular size becomes
\begin{eqnarray}
\label{ang}
\delta \phi = \frac{l}{d_L(z)} (1 + z)^2,
\end{eqnarray}
with $d_L$ defined above.

From the solutions represented by equation~(\ref{ab1}) we obtain for the
angular size,
\be
\label{ang1}
\ds \delta \phi_n\,=\, \frac{l}{a_0} \,
\frac{\left (\,1 \,+\, z\, \right )}
{sin \left [\frac{n}{1-n}\,\frac{1}{a_0\,H_0}
\left [\,1\,- \,\left (\,1\, +\, z \right )^{-\frac{1-n}{n}}\,\right ]\,\right ] },
\en
where $H_0 $ is  a parameter   defined by $\ds H_
0\,=\,\frac{n}{a_0^{1/n}}$.

\begin{figure}[ht]
\centerline{ \psfig{file=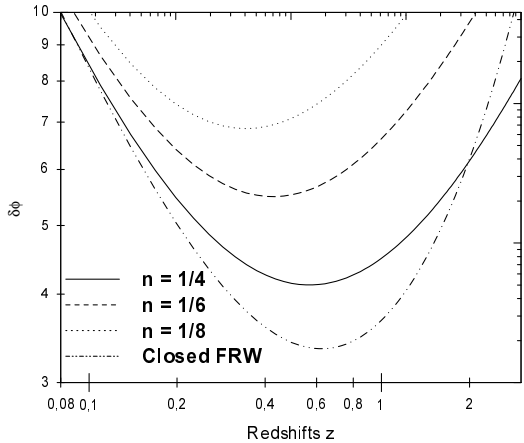,width=9cm,angle=-90} }
\caption{This plot shows the angular size (in unit of $l /a_0$) as
a function of the redshift $z$, for three different values of the
parameters $n$, $n = 1/4, 1/6, 1/8$. The   dash-dot-dot line
corresponds to the  isotropic closed FRW model, with a matter
component defined by the equation of state $\alpha = -1/3$.}
\label{fig1}
\end{figure}

Fig.~\ref{fig1} shows the angular size
as a function of the redshift in the range $0.05 \leq z \leq 2.80$
for three different values of the parameters $n$. Here, we have
used de value $a_0 H_0 = \sqrt{2/5}$. In this plot we have added
the graph of the angular size corresponding to the isotropic
closed FRW model, with a matter dominated by a quintessence
component defined by $\alpha = -1/3$. Notice that, for different
values of the parameter $n$, those curves  near to the value $n =
1/3$  become  closer to that corresponding of the isotropic FRW
model.  

\begin{figure}[ht]
\centerline{ \psfig{file=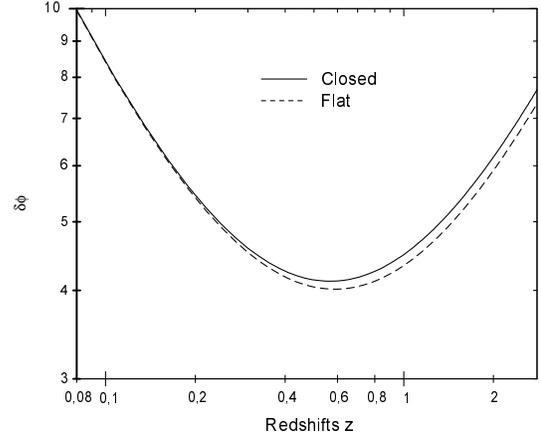,width=9cm,angle=-90} }
\caption{This plot shows the angular size (in unit of $l /a_0$) as
a function of the redshifts $z$,  for flat  and closed anisotropic
models. The parameters $n$ was chosen to be $n = 1/4$. Notice that
at low redshift the models become indistinguishable.} \label{fig2}
\end{figure}

In Fig.~\ref{fig2} we show the angular sizes as a function of the redshift $z$
for a flat and close anisotropic universe models. Notice that at low redshift
both curves become similar. We could distinguish them at $z \gtrsim 0.5 $.

\subsection{Cases $\eta'\,\neq\,0$}

As before, in this case we consider two different solutions. We start  by
describing a quasi-anisotropic  and an exponential growing solutions.

In the former case, we found as a possible solution $a(t) =
t^{2/3}$ and $\ds b(t) = t^{2/3}\, \left [1 + (t/t_0)^{-n}
\right]$, where $n$ and $t_0$ are two arbitrary constants. Notice
that at large time $b(t)$ approach to $a(t)$ and thus the universe
isotropizes. Thus, asymptotically the universe approaches to an
homogeneous isotropic flat universe which is filled by dust, i.e.,
$b(t) \sim a(t) = t^{2/3}$.

For this solution it is found that,
\be
\ds \rho_M = \frac{4}{3 \kappa t^2}
\left [ 1-\frac{n}{1+\left (\frac{t}{t_0}\right )^ n} \right],
\en
\be
\ds p_M = \frac{2 n (1-n)}{3 \kappa t^2
\left [ 1+\frac{n}{1+\left (\frac{t}{t_0}\right )^ n} \right]},
\en
and
\be
\ds \eta_{_{M}} = \frac{1}{2 n \kappa t}\left [ n \kappa (n-1)
+ t^{2/3} \left [ 1+\frac{n}{1+\left (\frac{t}{t_0}\right )^ n} \right] \right ].
\en
Notice that $\gamma$ becomes a time depending quantity in this case,
\be
\ds \gamma(t) = \frac{n}{2 (1-n)} \left [ 1-n +
\left ( \frac{t}{t_0}\right )^n \right]^{-1}.
\en
Notice also that $\gamma(t) \longrightarrow 0 $ for $t \longrightarrow \infty$, in
agreement with the remark described above.

The effective shear viscosity becomes $\ds \eta' = \frac{n-1}{2 \kappa t}$, and
in order to be positive the parameter $n$ should be bounded from below,
i.e., $n \geq 1$.

The corresponding solution for the scalar field is
\be
\ds Q(t) = \sqrt{\frac{6}{\kappa}} \left ( \frac{t}{t_0}\right )^{\frac{1}{3}}
\equiv Q_0 \left ( \frac{t}{t_0}\right )^{\frac{1}{3}},
\en
and the potential becomes
\be
\ds V(Q) = V_0 \left ( \frac{Q_0}{Q}\right )^4,
\en
where the constant $V_0$ is given by $V_0 = \frac{1}{9} Q_0^{-3}$.
This sort of solution was described in ref \cite{CrdeHe} where
scalar fields in FRW metric were studied.

A second possible solution is that in which the scale factor $a$
grows exponentially, i.e.,
\begin{eqnarray}
\label{a2}
a(t)= e^{Ht},
\end{eqnarray}
and
\begin{eqnarray}
\label{b2}
\ds b(t)=\, e^{-Ht/2} \, sin \, \left (\frac{3 H t}{2} \right ),
\end{eqnarray}
where   $H$  is a constant to be determined later on.

This solution corresponds to a universe filled with a viscous dust, since
$p_M = 0$  and $\eta' \neq 0$ at any time. The energy density and the
effective shear viscosity become
\be
\ds  \rho_{_{M}} =   3 H^2 cot \left( \frac{3 H t}{2} \right ).
\en
and
\be
\ds \eta'  = \frac{3 H}{2 \left[ cot \left (
\frac{3 H t}{2}\right ) - 1 \right]},
\en
respectively.

In order to have $\eta' \geq 0$ we must impose that $ \ds 0 \leq
\frac{3 H t}{2} \leq \frac{\pi}{4}$. This result in an age for the
universe  given by $\ds t_0\,=\,\frac{\pi}{6}\,H^{-1} $, which
could be used for fixing the value of $H$.

Notice that the solutions (\ref{a2}) and (\ref{b2}) give rise to
the following Hubble expansion rates,
\begin{eqnarray}
H_{1}=\frac{\dot{a}}{a}= H
\end{eqnarray}
and
\be
\ds H_{2}=\frac{\dot{b}}{b}= \frac{H}{2} \left[ 3 cot \left
( \frac{3 H t}{2}\right )- 1\right ],
\en
and thus the Hubble horizon related to the $\theta$-$\phi$ plane remains
constant.

The corresponding scalar field is found to evolve as
\begin{eqnarray}
Q(t)= \frac{1}{H} \sqrt{\frac{2}{3 \kappa}} \left[ e^{- H t_{0}}- e^{H t}
\right] + Q_{0},
\end{eqnarray}
where $Q_0$ is the value of $Q(t)$ at $t=t_0$.
The corresponding scalar potential $V(Q)$ becomes
\begin{eqnarray}
\ds V(Q)= \,V_0\,\left[ 1- \sqrt{\frac{3
\kappa}{2}} (Q-Q_{0}) \right],
\end{eqnarray}
where $V_0$ is a constant defined by $\ds V_0 = \frac{2}{3 \kappa} \, e^{- 2 H t_{0}} $.
From this expression we see that this potential decreases when $Q$ increases,
similar to the other case.



It is well known that the production of entropy could be related
to the anisotropy of the universe\cite{Mi,We}. In the following we
proceed to calculate this production in the cases described above.
In order to do this, we introduce the entropy current four-vector
$S^{\mu}$ as
\begin{equation}
S^{\mu} = n_{_{b}} k_{_{B}} \lambda u^{\mu},
\end{equation}
where as before $u^{\mu}$ represents the four-velocity, $n_{_{b}}$ the baryon
number density, $k_{_{B}}$ is the Boltzmann's constant  and $\lambda$ the
nondimensional entropy per baryon. It could be shown that\cite{MiThWe}
\be
\ds S^{\mu}\,\,_{; \mu} = \frac{2 \eta}{T} \sigma_{\mu \nu}\,\, \sigma^{\mu \nu},
\en
where in the first case we get
\be
\ds S^{\mu}\,\,_{; \mu} = \frac{2 n^2 (n-1)}{3 \kappa T t^3
\left[ 1+ (t/t_{0})^n  \right ]^2}.
\en
The left-hand side of this expression gives, in the comoving frame of reference
\be
\ds S^{\mu}\,\,_{; \mu} = k_{_{B}} n_{_{b}} \dot{\lambda},
\en
where we have used the conservation equation for baryon number,
$(n_{_{b}} u^\mu)_{; \mu} =0$.
Thus,  we get
\be
\label{entropia}
\ds \dot{\lambda} = \frac{2 n^2 (n-1)}{3 n_{b} k_{_{B}} T t^3 \left[ 1+ (t/t_{0})^n
\right ]^2}. 
\label{49}
\en
Note that this expression decrease when $t$ increase and becomes
zero for $t \longrightarrow \infty$, similar to $\eta_M$, $p_M$ and $\rho_M$.

From expression~(\ref{entropia}) evaluated at $t=t_{_{1000}}$
(equivalent to 1000 s) and $t=t_{rec}$ (time at recombination) we
get
\begin{eqnarray}
\frac{\dot{\lambda}_{_{1000}}}{\dot{\lambda}_{rec}}=
 \frac{n^{rec}_{b}}{n^{1000}_{b}}
\frac{T_{rec}}{T_{_{1000}}} \left (\frac{t_{rec}}{t_{_{1000}}} \right )^3 \left[
\frac{1+t^{n}_{rec}/t^n_{0}}{1+ t^{n}_{_{1000}}/t^n_{0}}.
\right]^2
\end{eqnarray}
With the data given in ref.~\cite{Brevik} and taking $n=2$, we obtain for
$t_{0}$ the value $t_{0} \approx 4 \times 10^{10}$ s. This value, together
with the age of the universe, $t_{c} \approx 5 \times 10^{17}$ s, allows us to
obtain the ratio between the shear, $\sigma$, and the scalar expansion,
$\theta$, given by
\begin{eqnarray}
\left( \frac{\sigma}{\theta}  \right)_{t_{c}}= \frac{n}{\sqrt{3} \, (2
(t_{c}/t_{0})^n + 2 - n)} \approx 5 \times 10^{-15},
\end{eqnarray}
which is inside of the bound expressed by COBE
measurements, that gives $\ds \left( \frac{\sigma}{\theta} \right)_{t_{c}} \leq
6.9 \times 10^{-10}$~\cite{Martinez,Takeshi}.

In the second case, and following a similar process we find that
\be
\ds \dot{\lambda} = \frac{9 H^3}{4 n_{b} k_{_{B}} \kappa T} \left[
cot \left ( \frac{3 H t}{2} \right )-1 \right]. 
\en
By using  the observational data specified above we get that
\be
\ds \frac{\dot{\lambda}_{_{1000}}}{\dot{\lambda}_{rec}}\,\simeq \,6\,\times\,10^{-25}.
\en
Thus, the generation of entropy has decreased more than $10^{25}$ times the value
at recombination during the period from $t_{rec}$ to $t \simeq 1000 s$.


\mbox{} \\ 
\section{Conclusions}

We have studied an anisotropic universe cosmological model described by the
metric~(\ref{metric}). We included in our model negative anisotropic pressures
motivated by quintessence cosmological scenarios. This component
was represented by a scalar field $Q$, whose equation of state was considered
to be given by $p_{_Q}\,=\, \alpha\,\rho_{_Q}$, where the parameter
$\alpha$ was considered to be equal to $-1/3$.

In order that our closed universe scenario could resemble a flat
model, we imposed the conditions  specified by equations
(\ref{cond1}) and (\ref{cond2}). Under these conditions, we have
determined, in different cases, explicit expressions for the
scalar potential $V(Q)$. In all these cases we have found that
these potential decrease as a function of the scalar field $Q$. In
this respect, would be interesting to study the cosmological
consequences that this sort of potential may have  during the
evolution of the universe. Specially, the influence that it
carried during the rapid expansion (inflation) that the universe
is believed to present at early time of its evolution.

In the cases in which the shear viscosity was vanished we have
determined the angular sizes for different values of the
parameters. Here, we found, similar to the isotropic case, that
our closed model looks similar to a flat model at low redshifts.

On the other hand, solutions in which the shear viscosity was not vanished,
we have determined the generation of entropy. Here, we have found that our results
agree with the bound imposed by the observational data.

\mbox{} \\
\section*{Acknowledgments}
MC was supported by COMICION NACIONAL DE CIENCIAS Y TECNOLOGIA through Grant
FONDECYT N$^0$ 1990601, also by Direcci\'{o}n de Promoci\'{o}n y Desarrollo
de la Universidad del B\'{\i}o-B\'{\i}o and in part by Dicyt (Universidad de
Santiago de Chile). SdC was supported from the COMICION NACIONAL DE CIENCIAS Y
TECNOLOGIA through Grant FONDECYT N$^0$ 1971157 and also from UCV-DGIP
123.744/99.
 


\begin{thebibliography}{2}

\bibitem{PeSchRi} S. Permutter {\it et al.}, Nature {\bf 391}, 51 (1998); {\it idem}
Report No. astro-ph/9812473; {\it idem}, Report No. astro-ph/9812133[Astrophy.
J.(to be published)]; B. Schmidt {\it et al.}, Astrophy.J. {\bf 507}, 46 (1998);
A. G. Riess {\it et al.}, Astrophy. J. {\bf 116}, 1009 (1998).

\bibitem{LiEfCo} C. Lineweaver, Astrophy. J. {\bf 505}  L69 (1998);
G. Efstathiou {\it et al.}, Mon. Not. R. Astro. Soc. {\bf 303}, 47 (1999);
K. Coble {\it et al.}, Report No. Astro-ph/9902195.

\bibitem{GuLiAl} A.H. Guth, Phys. Rev. D {\bf 23} (1981) 347;
A.D. Linde, Phys. Lett. B {\bf 108} (1982) 389;
A. Albrecht, P.J. Steinhardt, Phys. Rev. D {\bf 48} (1982) 1220;
A.D. Linde, Phys. Lett. B 129 (1983) 177.

\bibitem{Peetal} S. Perlmutter {\it et al.}, Phys. Rev. Lett. {\bf 83} (1999) 670.

\bibitem{ViSh} A. Vilenkin and P. Shellard, {\it Cosmic string and
other topological defects} (Cambridge University Press,
Cambridge, England, 1994)

\bibitem{Caetal} R.R. Caldwell, R. Dave, P. Steinhardt, Phys. Rev. Lett. {\bf 80}
(1998) 82.

\bibitem{Heetal} G. Huey, L. Wang, R. Dave, P.R. Caldwell, P. Steinhardt, Phys.
Rev. D {\bf 59} (1999) 063005.

\bibitem{Ag}A. N. Aguirre, Astrophy. J. {\bf 512}, L19 (1999); {\it idem}
astro-ph/9904319 (Astrophy.J. in press).



\bibitem{WhSc}M. White and D. Scott, Astrophy. J. {\bf 459}, 415 (1996).


\bibitem{KaTo} M. Kamionkowski and N. Toumbas, Phys. Rev.
Lett. {\bf 77 }, 587 (1996).


\bibitem{Da} R. L. Davis, Phys. Rev. D {\bf 35}, 3705 (1987);
Gen. Relativ. Gravit. {\bf 19}, 331 (1987).

\bibitem{Ko} E. W. Kolb, Astrophys.J. {\bf 344}, 543 (1989).

\bibitem{Ba1} J. D. Barrow, Phys. Rev. D {\bf 55}, 7451 (1997).

\bibitem{Ba2} J. D. Barrow and R. Maartens, Phys. Rev. D {\bf 59}, 043502 (1999).

\bibitem{Ze}Ya. B. Zel'dovicch, in {\it Physics of the Expanding Universe},
edited by M. Demianski (Springer, Berlin, 1979).

\bibitem{Mc}M. A. H. McCallum, in {\it General Relativity: An Einstein Centenary
Survey}, edited by S. W. Hawking and W. Israel (Cambridge: Cambridge
University Press, 1979).


\bibitem{KaSa} R. Kantowski and R. K. Sachs, J. Math. Phys. {\bf {7}}, 443 (1966).

\bibitem{MiCr} J. P. Mimoso and P. Crawford, Class. Quantum Grav. {\bf {10}}, 315 (1993).

\bibitem{BaDa} J. D. Barrow and M. P. Dabrowski, Phys. Rev. D {\bf{55}}, 630 (1997).

\bibitem{BySc} S. Byland and D. Scialom, Phys. Rev. D {\bf{57}}, 6065 (1998).

\bibitem{CrdeHe}N. Cruz, S. del Campo and R. Herrera,
Phys. Rev. D {\bf 58},123504 (1998).

\bibitem{Mi} C. W. Misner, Astrophy. J. {\bf 151}, 431 (1968).

\bibitem{We} S. Weinberg, Astrophy. J. 168 (1971) 175.

\bibitem{MiThWe} C. W. Misner, K. S. Thorner and J. A. Wheeler, {\it Gravitation}
(Freeman and Co., New York, 1973).

\bibitem{Brevik} I. Brevik, S.V. Pettersen, Phys. Rev. D 56 (1997) 3322.

\bibitem{Martinez} E. Martinez, J. L. Sanz, Astron. Astrophys. 300 (1995) 346.

\bibitem{Takeshi} T. Chiba, S. Mukohyama, T. Nakamura, Phys. Lett. B 408
(1997) 47.
\end{thebibliography}
\end{document}